\documentclass[11pt,a4paper]{article}
\usepackage{caltn,psfig}

\begin{document}

\docnum{IACHEC Report Series\#2}
\title{Summary of the 2014 IACHEC Meeting}
\author{David Burrows$^a$, Fabio Gastaldello$^b$, Catherine E.~Grant$^c$, Matteo Guainazzi$^d$, Kristin Madsen$^e$, \\
Eric Miller$^c$, Jukka Nevalainen$^f$, Paul P.~Plucinsky$^g$, Steve Sembay$^h$ \\
\\
($^a$Pennsylvania University, 
$^b$INAF - IASF Milano, \\
$^c$MIT Kavli Institute for Astrophysics and Space Research, \\
$^d$European Space Astronomy Centre of ESA, \\
$^e$Space Radiation Lab., California Institute of Technology, 
$^f$University of Tartu, \\
$^g$Harvard-Smithsonian Centre for Astrophysics, \\
$^h$Department of Physics
and Astronomy, University of Leicester)
\\
}
\date{\today}                   

\maketitle

\begin{center}
{\small
{\bf Abstract} \\
We present the main results of the 9th meeting of the International Astronomical Consortium for High Energy Calibration (IACHEC), held in Warrenton (Virginia) in May 2014.
Over 50 scientists
directly involved in the calibration of operational and 
future high-energy missions gathered during 3.5~days to discuss the status of the X-ray payloads inter-calibration, as well as possible ways to improve it. 
Sect.~2 of this Report summarises our current understanding of the energy-dependent inter-calibration status.
}
\end{center}

The International Astronomical Consortium for High Energy Calibration
(IACHEC)\footnote{{\tt http://web.mit.edu/iachec/}}
is a group dedicated to supporting the cross-calibration
environment of high energy astrophysics missions with the ultimate goal
of maximising their scientific return. Its members are drawn from
instrument teams, international and national space agencies and other
scientists with an interest in calibration in this area. Representatives
of over a dozen current and future missions regularly contribute to the
IACHEC activities. Support for the IACHEC in the form of travel costs
for the participating members is generously provided by the relevant
funding agencies.  In addition, the 2014 IACHEC meeting was financially
supported by the {\it Chandra} mission, the {\it Swift} mission, and NASA/Goddard
Space Flight Center.

IACHEC members cooperate within working groups to define calibration
standards and procedures. The scope of these groups is primarily a
practical one: a set of data and results (eventually published in
refereed journals) will be the outcome of a coordinated and standardised
analysis of reference sources (``high-energy standard candles'').
Past, present and future high-energy missions can use these results as a
calibration reference.

The IACHEC meets yearly to report on the progress of the working groups
and define the next year's activities. The inaugural IACHEC meeting
was held in 2006 on neutral ground in Iceland, but since then has been
hosted by local IACHEC members in Europe, the United States or Asia. In
2014 the 9th IACHEC meeting was hosted by the {\it Swift} XRT team at Penn
State and was held at the Airlie Centre in Warrenton, Virginia. It was
attended by 51 scientists from the US, UK, Germany, the Netherlands,
Italy, Finland, Denmark, China, India, and Japan, as well as by
officials from NASA and ESA.

The format of the IACHEC meetings includes both plenary sessions where
instrument calibration status, working group summaries and other topics
of interest are presented and parallel splinter sessions where working
groups meet to discuss results and use the opportunity for face-to-face
data analysis sessions.

This Report summarises the main results of the 9th meeting. It is
organised as follows: Sect.~\ref{sect:wgs} describes the main results discussed by
each of the Working Groups. Sect.~\ref{sect:status} summarises
the cross-calibration status. For more details, readers are referred to
the presentations collected at the IACHEC meeting web page\footnote{{\tt http://web.mit.edu/iachec/meetings/2014/index.html}}.

\section{Working Group reports}
\label{sect:wgs}

\subsection{CCD}

The CCD Working Group met in conjunction with the Backgrounds Working
Group due to substantial overlap in interested participants. As always,
the CCD Working Group provided a forum for cross-mission discussion and
comparison of CCD-specific modelling and calibration issues. For future
meetings we are considering broadening the scope to include related
devices such as Active Pixel Sensors and CdZTe detectors which share at
least some of the physical principles and event processing schemes with
CCD-based detectors. At IACHEC 2014, we heard from {\it Chandra}/ACIS,
XMM-Newton/EPIC, and ATHENA.

We started the session with three presentations on non-X-ray backgrounds.
Terry Gaetz described efforts to characterise the spatial structure of
the particle background on {\it Chandra}/ACIS.  Both front- and
back-illuminated devices have energy-dependent spatial features, at
least some of which are likely due to charge transfer inefficiency. 
Applying the very faint (VF) mode filter, found in {\tt acis\_process\_events},
reduces both the total background and the non-uniformity.  More details
on this work can be found on the {\it Chandra} web site{\footnote{{\tt http://cxc.harvard.edu/cal/Acis/Cal\_prods/bkgrnd/nonuniformity/acisbg.html}}.

K.~D.~Kuntz described his research comparing soft proton flares on XMM-Newton
and {\it Chandra}. XMM-Newton/EPIC-pn sees substantially more soft proton flaring
than {\it Chandra}/ACIS-S3, and the flares seen by XMM-Newton are stronger than those
seen by {\it Chandra}. Part of the reason is that {\it Chandra} does not observe as
much in the region just inside the magnetosheath which has the highest
incidence of flares. It is not obvious, though, why the response of
{\it Chandra} is lower than XMM-Newton to soft protons.

Lorenzo Natalucci presented work by Simone Lotti of GEANT4 simulations
of the particle background for an X-ray micro-calorimeter at L2 like the
proposed X-IFU on ATHENA. The simulations were validated by comparison
to {\it Suzaku} XRS data. A number of background reduction techniques were
evaluated, such as an anti-coincidence detector and changes to the
design geometry and shielding. A related paper will appear in SPIE
proceedings shortly (Lotti, et al. 2014).

The second half of the session was devoted to particular calibration and
modelling efforts for {\it Chandra} and XMM-Newton. Nick Durham summarised work
evaluating the temperature-dependent CTI correction in the {\it Chandra} data
processing pipeline which has become increasingly important as the
spacecraft ages and temperature deviations become more common. Terry
Gaetz discussed his work on improving the low-energy gain calibration
for {\it Chandra}/ACIS-S1.

Steve Sembay described the evolution of deep traps in the XMM-Newton/EPIC-MOS
CCDs.  These types of traps have also been detected on {\it Swift} XRT, which
uses the same CCDs as MOS run at warmer temperatures, but some of the
trap characteristics appear to be different.  The rate of generation of
these traps varies between MOS1 and MOS2.  There was interest from the
attendees in applying the same kind of search to {\it Suzaku}/XIS or {\it Chandra}
ACIS, where these types of traps have not yet been reported.

Finally, Norbert Schulz reported on his study of bright X-ray sources
observed with the {\it Chandra} HETG, and the spectral differences found
between the standard timed-event mode and continuous clocking (CC) mode.
 He has found that problems in bright HETG spectra are not due to
problems with the calibration of CC-mode, but the effects of dispersed
secondary images, such as the X-ray halo, which in CC-mode cannot be
spatially filtered.  These data require intensive additional modelling
and data reduction depending on the chosen configuration. His
description of the problem and recommendations for planning future
observations are found on the {\it Chandra} web pages{\footnote{{\tt http://cxc.harvard.edu/cal/Acis/Cal\_prods/ccmode/ccmode\_final\_doc03.pdf}}.

\subsection{Contamination}

Despite the best efforts in instrument design and construction, several
recent missions have suffered molecular contamination after launch
(Marshall et al., 2004, Koyama et al. 2007, O'Dell et al. 2013).  The introduction of
unknown absorbing material into the light path of an X-ray instrument not
only reduces the effective area at soft energies ($\leq$ 1 keV), but it
greatly complicates the calibration since the amount, composition, and
spatial distribution of this contaminant must be determined on orbit.
During previous IACHEC workshops, molecular contamination arose as a
serious calibration issue in several standard candle working groups
({\it e.g.}, Plucinsky et al., 2012, Kettula et al. 2013), and so at the 2013
meeting a new contamination working group was established to address three
broad topics:
\begin{enumerate} 
\item comparison of contamination among instruments and missions, including 
chemical composition,
time dependence,
spatial dependence,
temperature dependence, and
environmental dependence;
\item mitigation for current instruments, including
celestial monitoring targets,
effects on calibration and science results,
and ``bake-out'' procedures; and
\item mitigation for future instruments, including
design,
procurement,
ground procedures, 
testing and calibration, and 
on-orbit monitoring.
\end{enumerate}

With these topics in mind, the contamination working group held its
inaugural meeting at the 2014 IACHEC workshop, gathering 12 of 19 members
for two sessions.  Representatives from operating missions presented the
history and current status of contamination in their soft X-ray CCD
instruments.  Herman Marshall presented details about {\it Chandra} ACIS, using
the LETG gratings to probe absorption edges with more accuracy than CCD
resolution allows.  The accumulation rate of the ACIS contaminant had
decreased from 2000 to 2009 but has increased from 2009 until the present.
The chemical composition of the contaminant has varied during the mission,
possibly indicating that there is more than one source of the contaminant.
Doug Swartz presented models of the spatial migration of ACIS contaminant,
comparing low- and high-volatility components to explain the spatial
distribution observed.  Steven Sembay presented XMM-Newton, which shows
varying contamination among the instruments from the apparently
uncontaminated EPIC-pn, through EPIC-MOS1, EPIC-MOS2 and the RGS which show
increasing (although still modest) levels respectively.  Eric Miller showed
similar variation among the four {\it Suzaku}/XIS detectors, which became
contaminated very rapidly at the beginning of the mission but appear to be
recovering, with low-energy effective area increasing at recent times.
Andy Beardmore showed that the {\it Swift} XRT has little evidence for
contamination, although line ratios from recent observations of SNR 1E
0102.2$-$7219 are consistent with a contamination absorption profile.
The differences between these instruments, even among nearly identical
instruments on the same spacecraft, will be investigated as the working
group moves forward.

Two upcoming missions were represented as well.  Maurice Leutenegger
presented plans for contamination mitigation on the Astro-H SXS, which has
a multi-stage thermal and optical blocking filter system with integrated
heaters to reduce the probability of contamination in the light path of the
very cold micro-calorimeter.  Vadim Burwitz presented the design of the
eROSITA CCD instrument, with filters and cold traps to reduce
contamination, and summarised the mitigation strategies to reduce build-up
on the ground.

The large attendance at these sessions gave testimony to the concerns about
contamination in current and future mission.  Going forward, the three
topics above will continue to be addressed, culminating in publication of a
legacy white paper to detail shared lessons learned for instrument design,
ground mitigation, first light targets, and monitoring strategies.

\subsection{Galaxy Clusters}

Since clusters of galaxies are stable on human time-scales, they are useful for building up large cross-calibration data
bases of different X-ray missions
using non-simultaneous observations.  This is the main rationale of the activities of the Galaxy Cluster Working Group.

At the 9$^{th}$ IACHEC meeting, we discussed an extension of our previous work on the XMM-Newton/EPIC and {\it Chandra}/ACIS effective area cross-calibration
(Nevalainen et al. 2010)
by using a much larger sample (HIFLUGCS), and a new tool, the {\it stack residuals method}.
In a nutshell, this method consists of calculating an energy-dependent ratio between fluxes measured by pairs of instruments, by dividing the model calculated on the spectrum of one against the spectrum (data)
of the other.
If the method is applied to an homogeneous class of sources, the ratios are stacked together to increase the reliability of results, minimising the impact of the scatter of individual data sets from the mean.
Readers are referred to Kettula et al (2013), and Read et al. (2014) for a description of the algorithm, and of its
limitations.

The results of applying the stack residuals method on the HIFLUGCS sample are shown in Fig.~\ref{gerrit.fig} (Schellenberger et al. 2014). 
\begin{figure}[h]
\begin{center}
\psfig{file=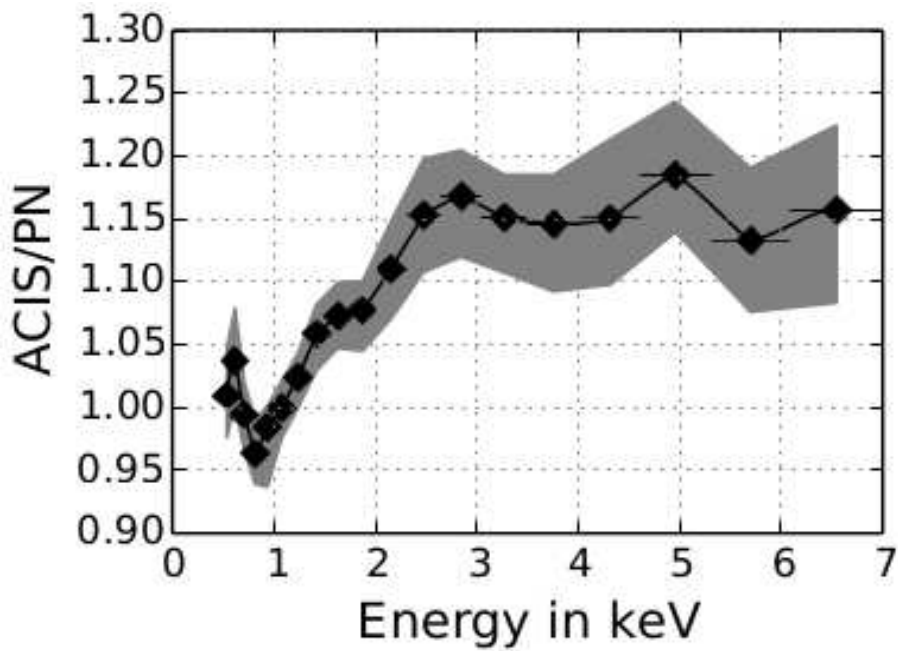,width=12.0cm}
\end{center}
\vspace{-0.5cm}
\caption{
ACIS/EPIC-pn stack residuals ratio using the HIFLUGCS sample (Schellenberger et al., 2014). The ratio is normalised to unity at 1.1~keV.
}
\label{gerrit.fig}
\end{figure}
We confirmed the previously published (Nevalainen et al., 2010) good agreement of the effective area shape
calibration above 2~keV. At lower energies the previously reported problems
remain. The relative uncertainties in the calibration of the effective area yield that ACIS measures systematically lower temperatures in the 0.5--7~keV energy band with respect to the EPIC cameras.
The better statistics of the HIFLUGCS sample allows a deeper study than in the original Nevalainen et al. (2010) paper,
and thus to derive the temperature dependence of the ACIS/EPIC temperature differences. Schellenberger et al. (2014) discuss the modification to the effective area calibration that would be
required to align the ACIS and EPIC spectral results\footnote{A tool implementing this correction
is provided at {\tt https://wikis.mit.edu/confluence/display/iachec/Data3}. A similar approach is discussed by Guainazzi et al. (2014), and implemented in SASv14.}.

Another activity discussed at the WG meeting is the "Multi Mission Study" aiming at comparing X-ray spectroscopic
results in the 0.5--10 keV band from 6 on-going 
and past X-ray missions (XMM-Newton, {\it Chandra}, {\it Swift}, {\it Suzaku}, NuSTAR and ROSAT) and 12 instruments. Preliminary
results for a sample of four clusters (A1795, A2029, 
Coma and PKS0745-19) indicate that, when compared to EPIC-pn, all other instruments yield higher flux above 2 keV; and
that all missions other than XMM-Newton yield lower 
flux below 2 keV (see Fig. ~\ref{mms.fig}). The data are not consistent with the hypothesis
whereby a single instrument has a significant bias 
in the calibration of the effective area while all the others are very accurately calibrated. 
The flux differences between different instruments vary up to $\pm$10\%,
\begin{figure}[h]
\begin{center}
\psfig{file=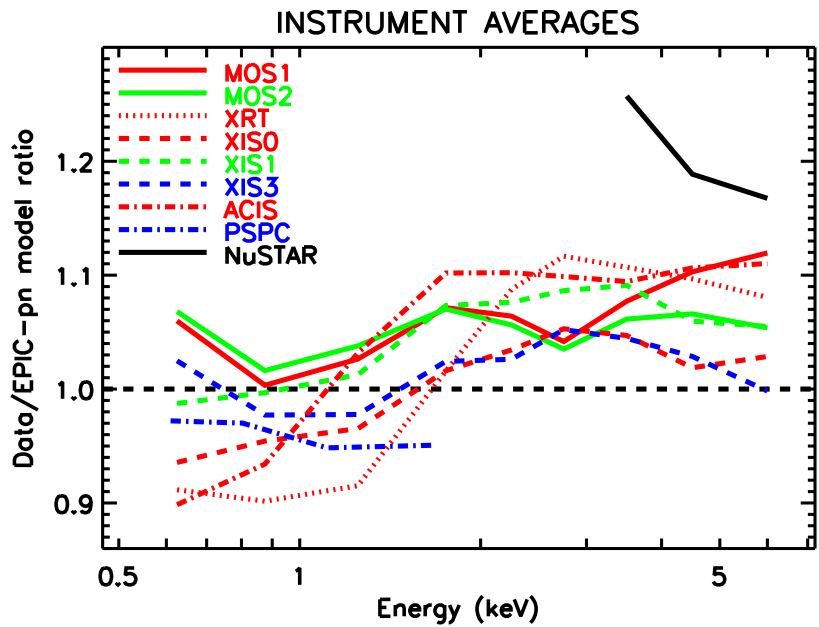,width=12.0cm}
\end{center}
\caption{
Cluster averages of the stack residuals for different instruments v.s. EPIC-pn as obtained in the
Multi Mission Project (Nevalainen et al. 2014 in prep.). The NuSTAR cross-normalisation is not representative of the level of discrepancy (a few percent) typically observed on point-like sources.
}
\label{mms.fig}
\end{figure}
except in the case of NuSTAR, which yields 
$\approx$20\% higher flux than EPIC-pn in the overlapping 4--7 keV band.
The higher NuSTAR cross-normalisation, based on a single observation at the center of Coma, has to do with aperture stop and single bounce reflection corrections that are not implemented
in the extended ARF mode of the current version of the NuSTAR analysis software. We expect the cross-normalisation to be dependent on the extraction region size and azimuthal location on the detector with respect the optical axis, with smaller areas close to the optical axis having better responses. These corrections are, however, fully taken into account for point sources, and the cross-normalisation factors for point sources are typically only on the order of a few percent. The extended ARF cross-normalisation numbers should therefore not be taken as a measure
of the actual NuSTAR effective area and vignetting functions for point sources, which apply to the vast majority of observations.

An alternative avenue for addressing the accuracy of effective area calibration is based on cluster
physics.
We studied the results for cluster total masses derived via the hydrostatic X-ray method and
gravitational lensing
published by other teams. While the X-ray masses derived with {\it Chandra}/ACIS are consistent with the masses derived via
gravitational lensing 
(Israel et al., 2014), the XMM-Newton/EPIC X-ray masses for another cluster sample are $\sim$30\% lower than  
the gravitational lensing masses (von den Linden et al., 2014). In principle, such a comparison could be used as an {\it absolute calibrator} of the effective area, overcoming the shortcomings inherent in the lack of X-ray
"standard candles" (Sembay et al., 2010). However, hydrostatic biases could lead to an overestimate in measurements of the cluster masses based on X-ray spectroscopy (Liu et al. 2009). For this reason, the
still existing astrophysical
uncertainties  prevents galaxy clusters from being used as absolute flux calibrators.

\subsection{Heritage}

A principal part of the scope of the IACHEC is the transmission of
expertise, good practice, and calibration-related methods from operational
missions to missions under development. While the path to instrument
calibration, as any other scientific undertaking, cannot be entirely foreseen,
there are significant gains for future missions in being able to avoid any 
errors and inefficiencies made in the past. These gains can be critical for
missions on tight budgets, or with shorter (predicted) operational lives.
Calibration plans are inevitably success-oriented due to the limited
time devoted to calibration, and time pressure during the development phase may
lead to losing the overall vision of long-term sustainable mission operations.
These shortcomings can be dearly paid for in the later phases of mission
operations. It is therefore crucially important to avoid following knowingly
wrong paths.

In order to systematise the IACHEC legacy,
a new "IACHEC Heritage Working Group" was founded at the 9th IACHEC meeting with the
following scopes:

\begin{itemize}

\item provide a platform for the discussion of experiences coming from operational
  missions
\item facilitate the usage of good practices for the management of pre- and
  post-flight calibration data and procedures, and the maintenance and
  propagation of systematic uncertainties (the latter task in strict
  collaboration with the "Systematic uncertainties" IACHEC Working Group)
\item document the best practices in analysing high-energy astronomical data as a
  reference for the whole scientific community
\item ensure the usage of homogeneous data analysis procedures across the IACHEC
  calibration and cross-calibration activities
\item consolidate and disseminate the experience of operational missions on the
  optimal calibration sources for each specific calibration goal
\end{itemize}

 The first tasks of this Working Group will be: a) to submit a synoptic paper
on in-flight calibration plans, that could inform the definition of the calibration
program for forthcoming X-ray missions; b) organise a workshop on photoelectric absorption models
and associated abundances and cross-section conventions with external experts.

\subsection{Thermal SNRs}

The Thermal SNR Working Group met with the following attendees
contributing in person: Steve Sembay (XMM-Newton/EPIC-MOS), Andy Beardmore ({\it Swift}/XRT),
Paul Plucinsky ({\it Chandra}/ACIS), Matteo Guainazzi and Martin 
Stuhlinger (XMM-Newton/EPIC), and Eric Miller ({\it Suzaku}/XIS), and the
following contributing remotely:   Frank Haberl (XMM-Newton/EPIC-pn) and Adam
Foster (APEC).  The group has used the standard IACHEC model for the 
Small Magellanic Cloud supernova remnant (SNR) 1E 0102.2-7219
(hereafter E0102) to understand time-dependent changes in their 
respective instrument responses (Plucinsky et al. 2008, 2012).  The 
E0102 model has been used to
identify issues with the ACIS contamination model and to verify
revised contamination models.  The {\it Chandra} X-ray Center released a new
ACIS contamination model on 9 July 2014 in CALDB release version
4.6.2, which was verified against E0102 observations from 2003 to 2014.
The E0102 model was also used to test the low-energy response files
for the MOS instruments.  Partially based on the E0102 results, the
MOS team released a contamination correction for the MOS instruments
in SAS 13.5 released on 9 December 2013.  The {\it Swift}/XRT team has used
E0102 to update the response matrices for both windowed timing (WT)
mode and photon counting (PC) mode data, resulting in significant
improvements at low energy over the course of the mission. The EPIC-pn
instrument appears to be the most stable instrument,
based on the consistency of the observed E0102 count rates over the
course of the XMM-Newton mission. Although there are differences at the few
percent level between E0102 observations acquired at different
positions on the EPIC-pn that are under investigation. 
The XIS team on {\it Suzaku} has continued to use E0102 to verify the XIS 
contamination model and make revisions as necessary, with an updated 
CALDB file released on 16 September 2013.  The bright emission lines of 
E0102 are also used to measure and verify the XIS gain and charge-transfer 
inefficiency (CTI) parameters at low energies.

Every year at the IACHEC, the various instrument teams present the line 
normalisations for the four major line complexes in the E0102 spectrum, 
specifically the O{\small VII}~triplet ($\sim 570$~eV), the 
O{\small VIII}~Ly$\alpha$ line ($654$~eV), the  Ne{\small IX}~triplet 
($\sim 915$~eV), and the Ne{\small X}~Ly$\alpha$) ($1022$~eV).  These 
normalisation are derived from representative observations in the
best-calibrated instrument modes and are based on the latest 
calibrations for the instrument.   The objective of this
exercise is to estimate the current agreement in the absolute
effective area of the instruments at these energies which are
particularly challenging to calibrate.  Fig.~\ref{fig3} plots the values of
\begin{figure}[h]
\begin{center}
\psfig{file=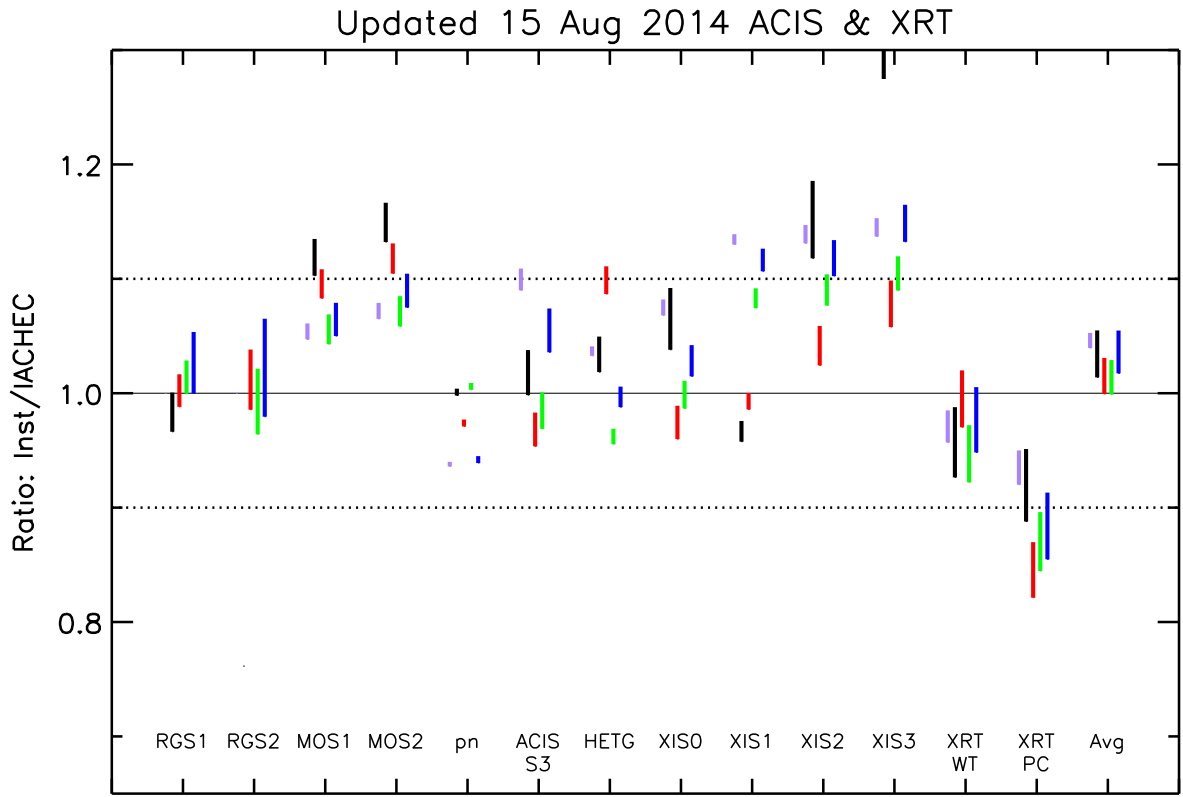,width=12.0cm,angle=90}
\end{center}
\vspace{-1.5cm}
\caption{
The fitted line normalisation for the various
instruments relative to the normalisations contained in the standard
IACHEC model for E0102. For each instrument, four or five lines are
plotted depending on if a global normalisation was used in the
fitting.  The global normalisation is indicated in purple, the 
O{\small VII}~triplet normalisation in black,  the 
O{\small VIII}~Ly$\alpha$ line normalisation in
red, the  Ne{\small IX}~triplet normalisation in green, 
and the Ne{\small X}~Ly$\alpha$) line normalisation in blue.
The dashed lines are drawn at $\pm10\%$ for reference.
}
\label{fig3}
\end{figure}
these normalisations with respect to the values in the standard IACHEC
model.  We do not claim that the values in the IACHEC model are
correct, we simply are using them as a convenient reference to compare
against.  41 of the 52 normalisations agree to within $\pm10\%$,
indicating the general level of agreement.  The RGS values agree well
with the standard values since the RGS data were relied upon to
develop the standard model.  The MOS1 and MOS2 data appear to be about
10\% high.  The EPIC-pn values range from 0--5\% below the standard
model. The ACIS normalisations range from $-5$ to $+5\%$ compared to
the standard model, with no clear energy dependence.  The XISs exhibit
different agreement for the different detectors.  The XIS0 and XIS1
agree to within 10\% but the XIS2 and XIS3 appear to be high by
$\sim10\%$.  The XRT WT mode data agree to better than 10\% but the
XRT PC mode data appear to be low by about 10\%.  The MOS and EPIC-pn
normalisations disagree by about 10\% for E0102, whereas they agree to
within 5\% for point sources on-axis (Read, Guainazzi, \& Sembay
2014). This apparent discrepancy is under investigation.

  The group also continued work on the development of a standard model
for the Large Magellanic Cloud (LMC) SNR N132D, the brightest SNR in
the LMC.  The standard model was developed based on the EPIC-pn and RGS
data. The other instrument teams compared this standard model to their
data and suggested additions of some weaker lines.  After these changes
have been made, the goal will be to use the N132D standard model to
compare the absolute effective areas in the band pass from 1.5 to 3.5
keV. Pileup is unfortunately an issue since N132D has such a high
surface brightness.  Only the EPIC-pn data in small window mode can be used
for this analysis since data in other modes with longer readout times
have significant pileup. The ACIS data in full-frame mode cannot be
used for this analysis since pileup is significant. We are exploring
the use of the 0th order data from an ACIS/HETG observation of N132D
to use for this analysis. The existing XIS data should not suffer 
pile-up and should be suitable for this analysis.
  
\section{Summary of the cross-calibration status}
\label{sect:status}

Fig.~\ref{fig4} represents a synopsis of cross-calibration measurements in the 0.1--10~keV energy band
\begin{figure}[h]
\begin{center}
\psfig{file=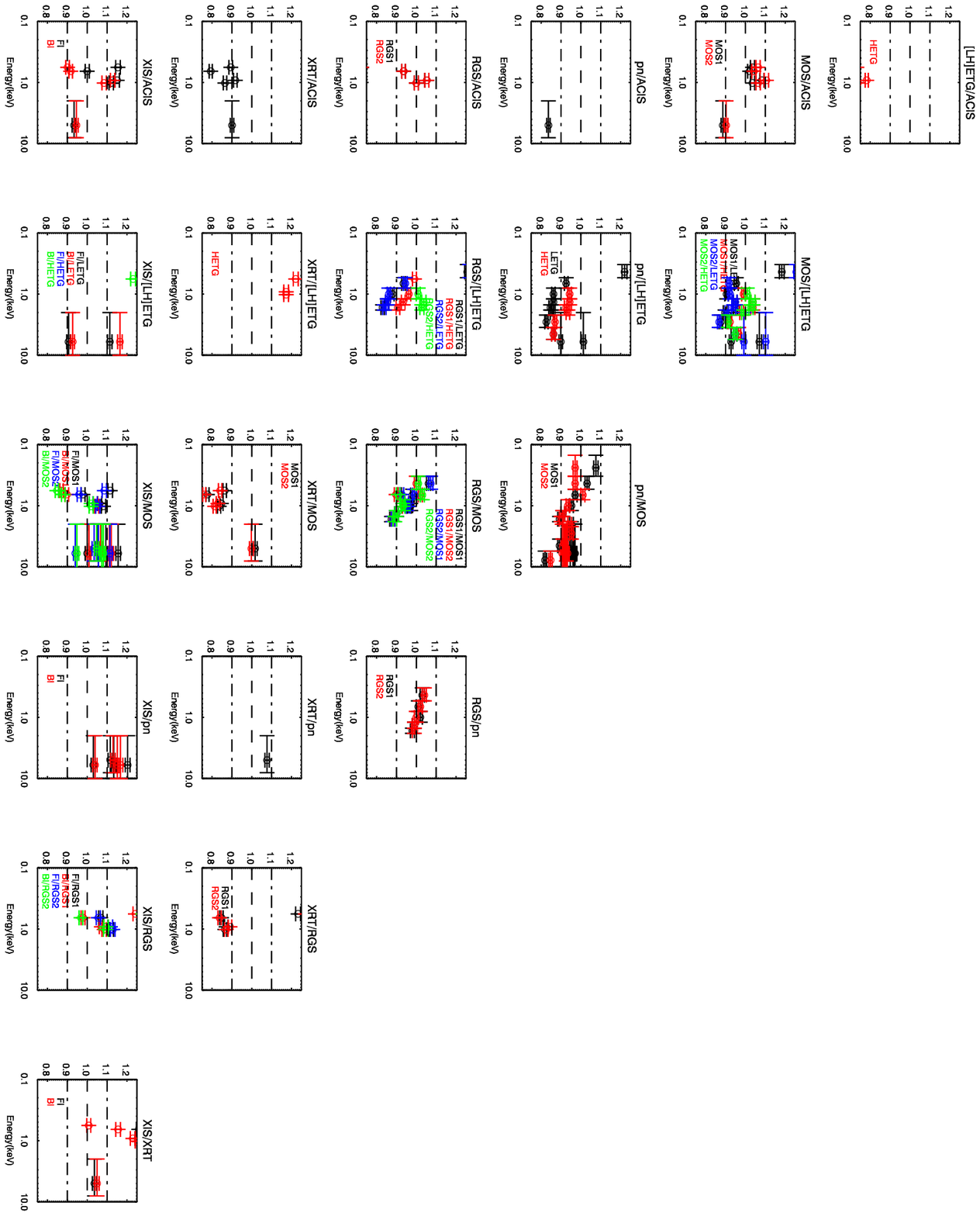,width=16.0cm}
\end{center}
\vspace{-1.5cm}
\caption{
Flux ratios as a function of energy in the 0.1--10~keV energy band for pairs of operational instruments compiled from IACHEC cross-calibration papers published in the last three years.
}
\label{fig4}
\end{figure}
recently published by the IACHEC (Tsujimoto et al., 2011, Ishida et al. 2011, Plucinsky et al. 2012; Smith \& Marshall in preparation). It must be stressed that these results
were published at different times, and therefore do {\it not} correspond to a homogeneous set of calibrations. Readers are warmly encouraged to refer to the aforementioned papers for a
discussion of the calibration sets used to reduce and analyse the data.

Fluxes measured by different instruments typically differ
within a dynamical range of 15\%.  In some cases where multiple measurements in the same energy bands are available a large scatter is observed 
(see, {\it e.g.}, the MOS/LETG, and the ratios involving the XIS). This indicates
that time-dependent calibration
effects may still significantly affect the cross-calibration status.

For a discussion of the cross-calibration status above 10~keV, interested readers are referred to the seminal paper by Tsujimoto et al.  (2011).

\section*{References\footnote{see {\tt http://web.mit.edu/iachec/papers/index.html} for a complete list of IACHEC papers}}

\noindent
Guainazzi M., et al., 2014, XMM-CAL-SRN-0321\footnote{{\tt http://xmm2.esac.esa.int/docs/documents/CAL-SRN-0321-1-2.pdf}}

\noindent
Kettula L., et al., 2013, A\&A, 552, A47

\noindent
Koyama K., et al., 2007, PASJ, 59, 23

\noindent
Ishida M., et al., 2011, PASJ, 63, 657

\noindent
Israel et al., 2014, A\&A, 564, 129

\noindent
Lau et al., 2009, ApJ, 705, 1129

\noindent
Lotti et al., 2014, Proc. SPIE, 9144

\noindent
Marshall H., et al., 2004, SPIE, 5165, 497

\noindent
Nevalainen J., et al., 2010, A\&A, 523, 22

\noindent
O'Dell S.~L., et al., 2013, SPIE, 8859

\noindent
Plucinsky P., et al., 2008, SPIE, 7011, 68

\noindent
Plucinsky P., et al., 2012, SPIE, 8443, 12

\noindent
Read A., et al., 2014, A\&A, 

\noindent
Schellenberger et al., 2014, A\&A

\noindent
Sembay S., et al., 2010, AIPC, 1248, 593 

\noindent
Tsujimoto M., et al., 2011, A\&A, 525, 25

\noindent
von den Linden et al., 2014, A\&A, submitted, (arXiv:1402.2670)

\end{document}